\documentclass[conference,10pt]{IEEEtran}

\usepackage{graphicx}
\usepackage{array}
\usepackage{color}
\usepackage{cite}
\usepackage{epstopdf}
\usepackage{float}
\usepackage{times,amssymb,amsmath}
\usepackage{lineno}

\begin{document}
\title{Interference Cancellation for UWA Random Access Data Packet Transmission} 
\author{Y.~Zakharov, L.~Shen, B.~Henson, N.~Morozs and P.~D.~Mitchell \\ ~\\
\small School of Physics, Engineering and Technologies, University of York, UK   
}
\markboth{IEEE Journal of Oceanic Engineering}%
{\quad}
\maketitle

\begin{abstract}
In underwater acoustic (UWA) random access communication networks with multiple users and data packet transmissions, the packet collisions 
are the main cause of the network performance degradation. The aim of this paper is to investigate interference cancellation (IC) techniques capable of resolving such collisions in a low-complexity modem with single-carrier modulation and single transducer. More specifically, in this modem, the IC is used at multiple stages of the receiver. Firstly, the IC is performed for cancelling the multipath interference to improve the equalization performance in comparison with the linear equalization and Rake combining. Secondly, the IC removes the interference from collided data packets within extracted signal segments after identifying the collisions. Finally, the IC is applied to the received baseband signal to improve the data packet detection. The modem performance is investigated in a lake experiment with intensive multipath channels. The experimental results demonstrate high detection performance of the proposed modem design and show that the proposed IC techniques can significantly improve the throughput of random access UWA networks. 
\end{abstract}
\IEEEpeerreviewmaketitle

\section{Introduction}

Modern terrestrial communication systems normally are multiuser. The multiuser communication underwater is also in great demand for multiple applications. The underwater acoustic (UWA) multiuser communication however is challenging due to the complicated propagation channel characterised by a limited frequency bandwidth, substantial multipath delay-Doppler spread of transmitted signals, and long propagation delays. Despite technical challenges associated with designing such systems, significant progress has been achieved in this area in the past 30~years. 

Multiple access control (MAC) protocols, e.g. such as the ALAN protocol in~\cite{brady1994adaptive}, are capable of efficiently handling channel resources in UWA multiuser communication systems. However, they require feedback channels and due to the large latency and low available bandwidth associated with transmission in UWA channels, there is a strong demand for communication systems without feedback channels~\cite{cho2011successive}. Besides, there is an interest to simplify the design of underwater nodes collecting and transmitting data, without performing a significantly more complicated task - detection of communication signals. In this paper, we consider scenarios with multiple nodes transmitting data packets asynchronously to a node (base station) equipped with a receiver. 

Many works on UWA multiuser detection consider multi-channel (multiple antenna) receivers benefiting from the antenna array gain and/or high degree of spatial diversity~\cite{stojanovic1996multichannel, gray1997multiuser, tsimenidis2001underwater, yeo2001performances, page2005coordinate, stojanovic2006multichannel, calvo2008efficient, cho2011successive, cho2012asynchronous, yin2021iterative, cuji2023joint, cuji2023transmit}. Such receivers are complicated due to the excessive hardware, algorithmic complexity, and high computational load inherent to multi-channel processing; moreover, difficulties of precise implementation of antenna arrays and their calibration and installation underwater add to the complexity and make such systems costly. Here we are interested in a modem with a single transducer. 

Many works consider code-division multiple access (CDMA) for UWA networks~\cite{tsimenidis2001underwater, stojanovic2005multiuser, page2005coordinate, stojanovic2006multichannel, calvo2008efficient}. CDMA systems typically rely on spread-spectrum signals for user separation and gaining extra benefits due to the processing gain. However, this gain is achieved at the cost of the reduced data rates. As an alternative to CDMA systems, the interleave division multiple access (IDMA) systems~\cite{ping2006interleave} are attractive since they are capable of user separation potentially without the reduction in data rates; the works~\cite{tsimenidis2011turbo, aliesawi2011iterative, jing2020new} demonstrate benefits of IDMA for multiuser synchronous UWA downlink, though using spread-spectrum signals. In this paper, we consider the IDMA uplink without spread-spectrum.

There have been especially recently published works on orthogonal frequency-division multiplexing (OFDM) for multiuser UWA communication~\cite{wang2013asynchronous, li2023multicarrier, cuji2023transmit}. OFDM systems have their benefits, such as the low-complexity signal processing performed in the frequency domain, and drawbacks, such as the requirement for high linearity of the power amplifier. Other signals have also been designed for multiuser UWA communications, such as a combination of the hyperbolically frequency modulated chirp with pseudo-noise spreading sequences~\cite{bernard2019multiuser, bernard2020multiuser}. In this paper, we only address single-carrier modulation, although with a receiver benefiting from the frequency-domain signal processing.

Relatively low-complexity multiuser UWA receivers can be designed using interference cancellation (IC) techniques. The paper~\cite{brady1994adaptive} proposes an IC technique for resolving packet collisions in single path channels.   
In~\cite{yeo2001performances}, the multiple-access interference (MAI) cancellation is based on recursive successive IC and adaptive decision-feedback equalizer with a single-path channel model to regenerate the intersymbol interference (ISI). The work~\cite{zhou2020parallel} proposes the use of the parallel interference cancellation. In the paper~\cite{yang2009interference}, the IC for CDMA transmission is based on a simplified channel estimation using the cross-correlation between the received signal and spreading sequence. The receiver in~\cite{cho2011successive}, operating with an antenna array, cancels the MAI and ISI separately, assuming the channel estimates are available. The IC is essential for the IDMA, and the works~\cite{tsimenidis2011turbo, aliesawi2011iterative} consider IC designs for the synchronous downlink. In~\cite{yin2021iterative}, it is shown that with the IDMA and a long vertical receive antenna array, the IC based receiver allows up to nine user signals to be successfully decoded.     

The IC is based on channel estimation and equalization. The channel estimation is one of the most computationally demanding operations in the receiver, while its accuracy is essential for the receiver performance~\cite{stojanovic2005multiuser, page2005coordinate, calvo2008efficient}. Therefore exploiting any \textit{a priori} information on the channel, such as its sparsity~\cite{li2007estimation, huang2010comparison, berger2010sparse, qi2011underwater, junejo2018sparse}, and reusing tentatively decoded data as extra pilot symbols~\cite{cho2012asynchronous, yin2021iterative} is important to achieve a high estimation accuracy. These approaches are used in the modem presented in this paper.  

For the equalization, such techniques as the Rake combining (also known as the passive time reversal), linear equalization and decision-feedback equalization are most often used in practice. The Rake combiner performs the maximal-ratio combining (MRC) of multipath signal components and thus it provides the highest signal-to-noise ratio (SNR) at its output. However, such a combining ignores the multipath interference. Therefore, it is mainly acceptable in systems with spread-spectrum signals and multiple antenna systems, capable of operating at signal powers significantly lower than the noise and/or interference power, and often in combination with other equalization schemes~\cite{song2007multiuser, tsimenidis2011turbo}. Note that spread-spectrum signals have a low spectral efficiency, i.e., the transmission data rate is low. In high data rate systems, the classical Rake combiner is inefficient due to the multipath interference. Linear equalizers, such as the zero-forcing equalizer and minimum mean-square error (MMSE) equalizer, can efficiently deal with the ISI, but their performance might suffer from the noise enhancement. Decision-feedback equalizers are popular in UWA multiuser communication receivers due to their relatively low complexity and good performance~\cite{tsimenidis2001underwater, stojanovic2006multichannel, aliesawi2011iterative}. In our receiver, we present a merger of the Rake combiner and DFE, which was shown to result in a very good  performance.    

An important stage of signal processing in the receiver is the data packet detection. In application to the CDMA packet transmission, this was addressed in~\cite{stojanovic2005multiuser}. In our paper, we also address this problem and show how this stage can benefit from the IC.

The aim of many UWA multiuser modem designs is to try ``to bridge the gap between affordable computational load and acceptable performance"~\cite{calvo2008efficient}. In this paper, we also address this aim. In~\cite{zakharov2019data}, we proposed a low-complexity design of a single-user single-transducer modem. To use this modem in a network, MAC protocols can be designed. However, MAC typically 
involves exchanging control signalling among the nodes, thus making the whole system complicated and possessing large latencies. There is a need for a simple modem, with a single hydrophone, that can deal with multiple nodes 
equipped with transmitters only in a random access network. More specifically, the aim of this paper is to propose and investigate IC techniques capable of resolving collisions in a single-carrier modem design similar to that in~\cite{zakharov2019data} with data packet transmission in UWA random access networks. The contributions of the paper are as follows. 

\begin{itemize}
\item We propose a general structure of the single-carrier modem that can efficiently deal with data packet collisions in a random access network. This modem is a further development of the design presented in~\cite{zakharov2019data}. 
\item We propose an IC-Rake multipath combiner that by exploiting turbo-iterations in the receiver outperforms the Rake combiner and linear equalizer. Part of this material has previously been presented in the UACE-2023  conference~\cite{Shen2023_Rake}. Here, we present a more comprehensive description of the technique and show its relation to  the Rake combiner and linear equalizer. 
\item We propose IC techniques for resolving collisions in detected data packets and for improving the packet detection performance. 
\item We investigate the performance of the proposed designs in a lake experiment. 
\end{itemize}

The paper is organized as follows. Section~\ref{Sec:Packet} introduces the proposed structure of the transmitted data packet and describes the channel model. The receiver design is presented in Section~\ref{Sec:Modem-Design}. Section~\ref{Sec:Numerical-Sea} describes the lake experiment and presents its results. Section~\ref{Sec:Conclusions} concludes the paper.

\section{Transmitted data packet and channel model \label{Sec:Packet} }

In a data packet with the single-carrier modulation, the transmitted signal $y(t)$ represented by its samples $y(i)$ at a sampling frequency $f_s$ is given by 
\begin{eqnarray}
y(i) = \Re \left\{ e^{j 2 \pi (f_c / f_s)i} \sum_{n = 0}^{N-1} s(n) g\left(i-n \frac{f_s}{F_d} \right)    \right\} ,
\label{Eq:Transmitted-Signal}
\end{eqnarray}
where $s(n)$ is a sequence of superimposed data and pilot symbols, $N$ the number of symbols in the data packet,  $f_c$ the carrier frequency, $F_d$ the symbol rate, $\Re \{ \cdot \}$ denotes the real part of a complex number, and $j = \sqrt{-1}$; $g(i)$ is a pulse-shaping waveform, specifically the impulse response of the root raised cosine (RRC) filter with a roll-off factor~$\alpha$~\cite{proakis2008digital}. The superimposed symbols are given by
\begin{eqnarray}
s(n) = p(n) + j d(n) , \ \ n = 0, \ldots , N-1 , 
\label{Eq:Superimposed-Symbols}
\end{eqnarray}
where $p(n)$ and $d(n)$ are pilot and data symbols, respectively.  In the receiver, the pilot symbols $p(n)$ are used for the packet detection, user identification, Doppler estimation, time synchronization, channel estimation, and estimation of the signal-to-noise ratio (SNR). The pilot symbol sequence, which is user-specific, is a pseudo-noise sequence of length $N$, and $p(n)$ and $d(n)$ are BPSK symbols. The data symbols $d(n)$ are obtained by coding a data message, including a cyclic redundancy check (CRC), and applying to the coded sequence a user-specific interleaver, thus implementing the IDMA.   

Benefits of using time-frequency-energy resources in the superimposed transmission are discussed in detail in~\cite{ghogho2005channel, tugnait2006superimposed, he2008doubly, zakharov2015ofdm, zakharov2019data}. This approach however requires the self-interference from the data to be cancelled when estimating the channel, as well as the self-interference from the pilot when demodulating the data. The self-interference can be efficiently dealt with using turbo-iterations, which has been shown in applications to UWA single-carrier~\cite{zakharov2019data} and multi-carrier~\cite{zakharov2015ofdm} transmission.

\begin{table}[]
\caption{Example of data packet parameters \label{Table:Data_packet}}
\begin{center}
\begin{tabular}{lll}
\hline 
 Parameter:   &   Value:     \\
 \hline  
  Carrier frequency & $f_c = 32$~kHz      \\
 \hline      
  Sampling frequency & $f_s = 192$~kHz     \\
 \hline      
  Symbol rate & $F_d = 6000$~symbols/s     \\  
 \hline  
  Number of data symbols  & $N = 600$~symbols \\
 \hline  
  Pilot sequence & User-specific $N$-long segment\\
   & of a Gold sequence    \\
 \hline          
  Data packet length  & $100$~ms     \\
 \hline  
  Code rate   & $1/3$    \\
 \hline      
  Packet load  & $200$~bits     \\
  \hline
  CRC length  & $16$~bits     \\
 \hline  
  Code polynomials   & $(557, 663, 711)$ in octal  &    \\
 \hline      
  Interleaver & User-specific random interleaver \\
 \hline          
  Roll-off factor & $\alpha = 0.2$   \\
 \hline          
  RRC filter length & $L_\text{RRC} = 160$   \\
 \hline          
\end{tabular}
\end{center}
\end{table}
To ease understanding of further material in this paper, Table~\ref{Table:Data_packet} presents parameters of an example data packet. These parameters were used in the lake experiment described in Section~\ref{Sec:Numerical-Sea}.

The UWA channel is modelled as a time-variant linear system with an impulse response $h(t,\tau)$ that describes the multipath and Doppler spread. The noise-free signal at the receiver input is given by
\begin{equation}
x(t)=\int^{\infty}_{-\infty} h(t,\tau) y(t - \tau)d\tau ,
\label{Eq:Convolution}
\end{equation}
and the received signal is 
\begin{equation}
r(t) = x(t) + n(t) ,
\label{Eq:Received-Signal}
\end{equation}
where $n(t)$ is a noise signal. 

\begin{figure}
\centerline{\includegraphics[scale=0.6]{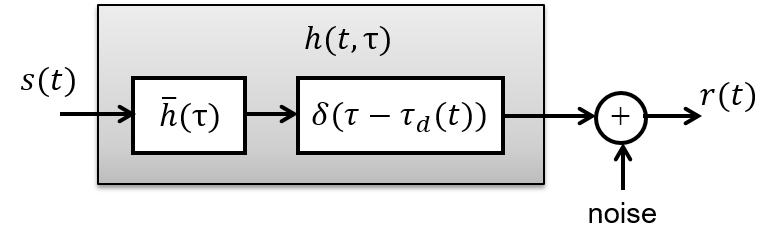}}
\caption{Representation of the channel as a time-variant filter. }
\label{Fig:Channel_Model}
\end{figure}
Over the duration of data packet that we consider, the UWA channel can be represented using a time-invariant filter with the impulse response $\bar{h}(\tau)$, modelling the multipath channel spread, and time-varying delay $\tau_d(t)$, describing the Doppler effect, as shown in Fig.~\ref{Fig:Channel_Model}. The delay $\tau_d(t)$ is normally equalized in the receiver via resampling, thus introducing a compensating delay $-\tau_d(t)$ into the received signal. The time-invariant component $\bar{h}(\tau)$ defined by a set of multipath delays and amplitudes is equalized using an inverse filter or Rake filter matched to the impulse response $\bar{h}(\tau)$. 
For simplicity, we will model $\tau_d(t)$ as $\tau_d(t) = a_0 + a_1 t$~\cite{mason2008detection, berger2010sparse, yerramalli2011optimal},
where $a_0$ is a constant delay, associated with the time difference between the transmitter and receiver; and with perfect time synchronization, $a_0 = 0$. The parameter $a_1$ is associated with a velocity $v$ between the transmitter and receiver, $a_1 = 1 + v/c$, where $c$ is the sound speed (e.g., $c = 1500$~m/s).

\section{Receiver 
\label{Sec:Modem-Design}}

In this section, we first describe the general structure of the receiver and main signal processing techniques used in the receiver (subsection~\ref{Subsec:Receiver-General-Structure}). We then describe the data packet detection in subsection~\ref{Subsec:Data-Packet-Detection}. The sparse channel estimator is revised in subsection~\ref{Subsec:Channel-Estimation}. The IC-Rake combiner is introduced in subsection~\ref{Subsec:IC-Rake-combiner}. Data packet regeneration and IC techniques for resolving data packet collisions and improving the data packet detection are presented in subsections~\ref{Subsec:Data-Packet-Recovery} and ~\ref{Subsec:IC-Data-Packet-Collisions}, respectively.

\subsection{General structure of the receiver \label{Subsec:Receiver-General-Structure}}

\begin{figure}
\centering
\includegraphics[width=7cm]{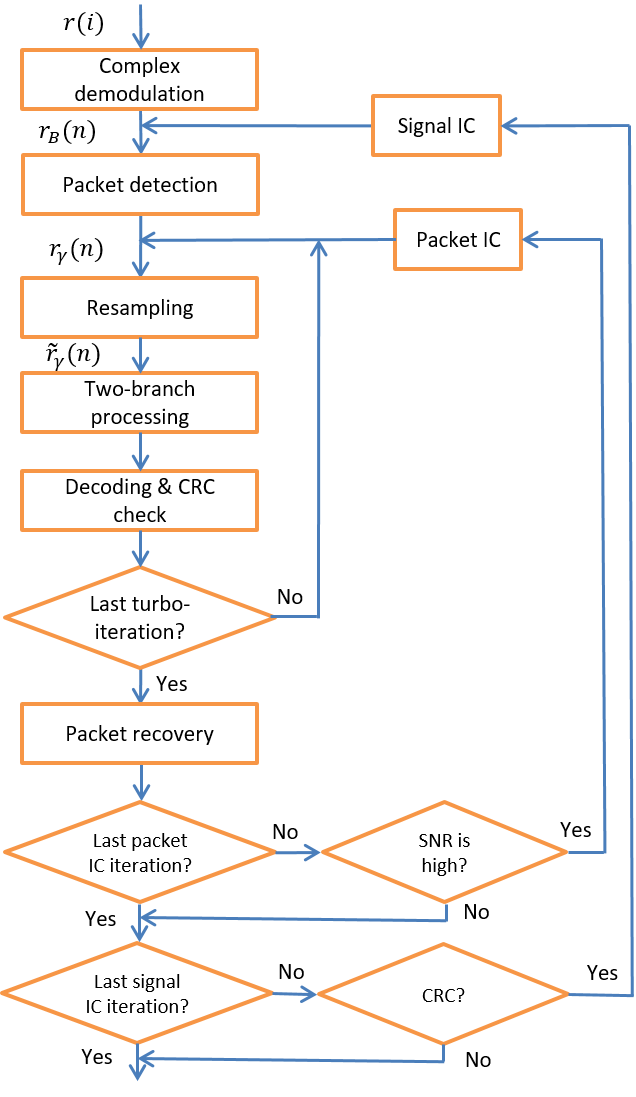}
\caption{General structure of the receiver. The received signal $r(t)$ is represented by samples $r(i)$ at the high sampling frequency $f_s$; the baseband signal is represented by samples $r_\text{B}(n)$ at the lower sampling frequency $F_{ds}$.}
\label{Fig:ICMUD-Receiver}
\end{figure}
The general structure of the receiver is shown in Fig.~\ref{Fig:ICMUD-Receiver}. The signal processing is first performed at a high sampling rate~$f_s$ chosen to avoid distortions due to the analog-to-digital conversion of the received passband signal: $f_s > 2f_c + (1+\alpha)F_d$. 
The complex demodulation transforms the passband signal into a baseband signal with overall bandwidth $(1+\alpha)F_d$ at a lower sampling rate $F_{ds}$, which, in our design, is $F_{ds} = 2 F_d$.  
The input signal $r(t)$ to the receiver is represented by real-valued samples $r(i)$ at the sampling rate $f_s$. The baseband signal samples are computed as: 
\begin{eqnarray}
r_\text{B}(n) =   \sum_{i = -L_\text{RRC}}^{L_\text{RRC}}  g(i) r(i- n K ) e^{-j 2 \pi (f_c/f_s) (i-nK)} ,
\label{Eq:Complex-Demodulation}
\end{eqnarray}
where $K = f_s / F_{ds}$ is the downsampling factor and $n$ is the time index at the sampling rate $F_{ds}$. 

The packet detection is based on a sliding-window monitoring of the received baseband signal with the aim to detect a ($\gamma$th) data packet by comparing the received signal with user specific pilot sequences, thus identifying the user, and to estimate the packet arrival time (timestamp) and Doppler parameters. The $\gamma$'s timestamp is used for windowing a signal segment $r_\gamma(n)$, $n = 0,\ldots,2N_\text{FFT}-1$, potentially containing a data packet, for further processing in the turbo iterations; the segment length $2N_\text{FFT}$ should be longer than the transmitted packet length ($N_\text{FFT} \geq  N$) and long enough to cover the received packet with its reverberation tail. A more detailed description of the packet detection is presented in subsection~\ref{Subsec:Data-Packet-Detection}. 

The signal segment $r_\gamma(n)$ is resampled according to the Doppler estimate: $\hat{\tau}_d(t) = \hat{a}_1 t$, assuming that the extracted data packet segment is symmetrical around $t = 0$. At the first turbo iteration, the estimate $\hat{a}_1$ obtained during the packet detection is used for the Doppler correction, i.e., the signal resampling and its Doppler frequency shift. At further iterations, these estimates are refined by the fine Doppler estimator. The resampled signal is denoted as $\tilde{r}_\gamma(n)$.

\begin{figure}
\centering
\includegraphics[width=8cm]{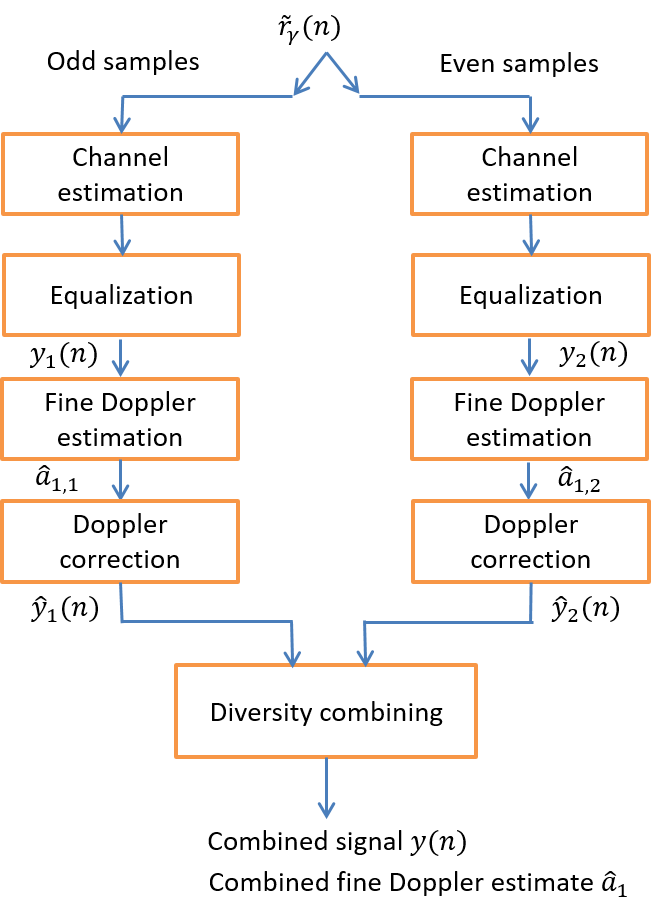}
\caption{Branch processing.}
\label{Fig:ICMUD-Branch processing}
\end{figure}
The Doppler corrected signal segment sampled at the  double symbol rate $F_{ds}$ is then split into two segments each with sampling rate equal to  the symbol rate $F_d$, with odd and even samples of $\tilde{r}_\gamma(n)$, respectively, and each segment is processed in a branch with the channel estimation and equalization as shown in Fig.~\ref{Fig:ICMUD-Branch processing}. 

Since UWA channel impulse responses are often sparse, channel estimators exploiting the sparsity are popular and useful in UWA modems~\cite{li2007estimation, huang2010comparison, berger2010sparse, qi2011underwater, junejo2018sparse}. In~\cite{zakharov2019data}, a low-complexity sparse channel estimator was presented, which was shown to demonstrate an estimation accuracy high enough to provide reliable demodulation of data packets in real experiments. We use this channel estimator in our receiver. Since the channel estimation is an essential technique for the IC, we revise the channel estimator from~\cite{zakharov2019data} in subsection~\ref{Subsec:Channel-Estimation}.  

The channel estimates are used in an equalizer, which can be the linear equalizer or the Rake combiner. In subsection~\ref{Subsec:IC-Rake-combiner}, we present another equalizer, the IC-Rake combiner, which benefits from data estimates obtained at the previous turbo iteration to subtract the ISI and from MRC combining the multipath signal components. The channel estimation and equalization are performed in the frequency domain to reduce the complexity. 

In each branch, the pilot part of the equalized symbols (their real part) is used to refine the Doppler estimate. This is based on the dichotomous frequency estimation~\cite{zakharov1999frequency, zakharov2001dft} as detailed in~\cite{zakharov2019data}. 
The fine Doppler estimate is applied to equalized symbols $y_b(n)$, $b = 1, 2$, $n = 0, \ldots, N-1$, for reducing the Doppler residual still present in the signal after the resampling.   

The Doppler corrected symbols $\hat{y}_b(n)$ are combined according to  
\begin{equation}
y(n) = w_1 \hat{y}_1(n) + w_2 \hat{y}_2(n) . 
\label{Eq:Diversity-Combining-Data}
\end{equation}
The weights $w_1$ and $w_2$ ($0 \leq w_b \leq 1$) for the (maximal-ratio) combining are computed based on SNR estimates in the two branches, 
\begin{equation}
\text{SNR}_b = \frac{1}{N} \sum_{n = 0}^{N-1} \left[  \Re \{ \hat{y}_b(n) \} - p(n)    \right]^2 , \ \ b = 1, 2.
\label{Eq:SNR-Estimate}
\end{equation}
The weights are then calculated as: 
\begin{equation}
w_b = \frac{\sqrt{\text{SNR}_b}} {\sqrt{\text{SNR}_1} + \sqrt{\text{SNR}_2}}.    
\label{Eq:w-Estimate}
\end{equation}
The estimate in~(\ref{Eq:SNR-Estimate}) is based on the fact that, in the case of perfect equalization, the real part of the signal $\hat{y}_b(n)$ should only contain the pilot signal $p(n)$ and any difference from $p(n)$ can be treated as noise. 

The fine Doppler estimates found in the two branches are also combined with the same weights: 
\begin{eqnarray}
 \hat{a}_1 = w_1 \hat{a}_{1,1} + w_2 \hat{a}_{1,2},
\label{Eq:Diversity-Combining-Doppler Estimates}
\end{eqnarray}
where $\hat{a}_{1,1}$ and $\hat{a}_{1,2}$ are estimates of the Doppler parameter $a_1$ in the two branches, respectively. 
The combined Doppler estimate $\hat{a}_1$ from~(\ref{Eq:Diversity-Combining-Doppler Estimates}) is used for the Doppler correction in the next turbo-iteration.

The combined signal is deinterleaved according to the identified user, decoded and the CRC check is performed. The decoded signal in the last turbo iteration is used to regenerate the received data packet as described in subsection~\ref{Subsec:Data-Packet-Recovery}. The recovered signal is used in the packet IC iterations, to remove the interference from overlapping packets being detected within the same signal segment; the collisions are identified according to the packet timestamps. The recovered signals are also used in signal IC iterations, where IC is applied to the original baseband signal before the packet detection.  The former iterations improve the receiver decoding performance on detected packets. However, since the packet overlap also affects the packet detection, the signal IC iterations are found useful for improving the packet detection performance. The IC iterations are detailed in subsection~\ref{Subsec:IC-Data-Packet-Collisions}.

\subsection{Data packet detection \label{Subsec:Data-Packet-Detection}}

This processing is based on computation of the cross-ambiguity function (CAF) between symbol-rate samples of an $N$-length segment $r_q(k) = r_\text{B}(2k - 2qQ)$, $k = 0, \ldots, N-1$, of the baseband signal $r_\text{B}(n)$, and user pilot signals $p_u(n)$, $u = 1,\ldots,U$, for all $U$ users on a grid of delay $\tau$ and Doppler frequency shift $F$: 
\begin{eqnarray}
A(u, q, \tau, F) = \sum_{n = 0}^{N-1}  
r_q^\ast(n) p_\text{u}\left(n - \tau \right) e^{-j 2 \pi (F/F_d) n} .   
\label{Eq:CAF}
\end{eqnarray}
The computations are repeated with a time step $Q$ of a half length of the data packet, $Q = N/2 = 300$ in our example, and $q$ is an index, which is incremented for every processed segment. In the CAF method, the number of processing branches $2N_d + 1$ is defined by the maximum speed $v_M$ of the receiver with respect to the transmitter. More specifically, $N_d = (v_M / c) f_c / \Delta_f$, where $\Delta_f$ is a Doppler frequency step (typically chosen as $\Delta_f = 1 / \Theta$), and $\Theta$ is the signal duration; $\Theta = 0.1$~s in our design. 
Let 
\begin{eqnarray}
I(u,q) &=& \max_{\tau, F} |A(u,q,\tau, F)|^2 .
\label{Eq:Max-A}
\end{eqnarray}

For the data packet detection, the following test has been found useful. If the segment $r_q(n)$ contains a substantial part of a data packet, then $I(u,q)$ should be high compared with $I(u,q-3)$ and $I(u,q+3)$, which will correspond to segments unlikely overlapping with data packets of the same user. The test 
\begin{eqnarray}
I(u,q) > I(u,q-3) + I(u,q+3)
\label{Eq:Test-1}
\end{eqnarray}
allows to check this condition. 

Another test checks that there is a clear peak in the~CAF, specifically 
\begin{eqnarray}
I(u,q) >  \frac{\nu}{2 N_d N_\text{FFT}} \sum_{\tau} \sum_{F, F \neq F_\text{max}} |A(u, q, \tau, F)|^2,
\label{Eq:Test-2}
\end{eqnarray}
where the parameter $\nu$ in our design is set to $\nu = 5$, and $F_\text{max}$ is found as 
\begin{eqnarray}
\{\tau_\text{max} , F_\text{max} \} = \arg \max_{\tau, F} |A(u, q, \tau, F)|^2 .
\label{Eq:A-tau-Doppler}
\end{eqnarray}

After finding $q$ for which~(\ref{Eq:Test-1}) and (\ref{Eq:Test-2}) are satisfied, the delay estimate $\tau_\text{max}$ from~(\ref{Eq:A-tau-Doppler}) is used to extract a signal segment $r_\gamma(n) = r_\text{B}(n - \tau_\gamma)$, $n = 0, \ldots, 2N_\text{FFT}-1$, for further baseband processing, where $\tau_\gamma = 2qQ - \tau_\text{max}$ is the timestamp used in the IC processing, and $\gamma$ is an index incremented after every new detected packet. The segment length $2N_\text{FFT}$ should be long enough to accommodate a received data packet of length $2N$ with its reverberation tail due to the multipath propagation. Such a segment can be considered as the channel output due to a zero-padded transmission of length $2N_\text{FFT}$. This allows the signal processing to be based on the Fast Fourier Transform (FFT) of the size $N_\text{FFT}$; in our example, $N_\text{FFT} = 1024$. The value $F_\text{max}$ is used to compute the initial Doppler parameter estimate $\hat{a}_1 = 1 + F_\text{max} / f_c$, further used for the Doppler correction of the received signal.

\begin{figure}
\centering
\includegraphics[width=9.5cm]{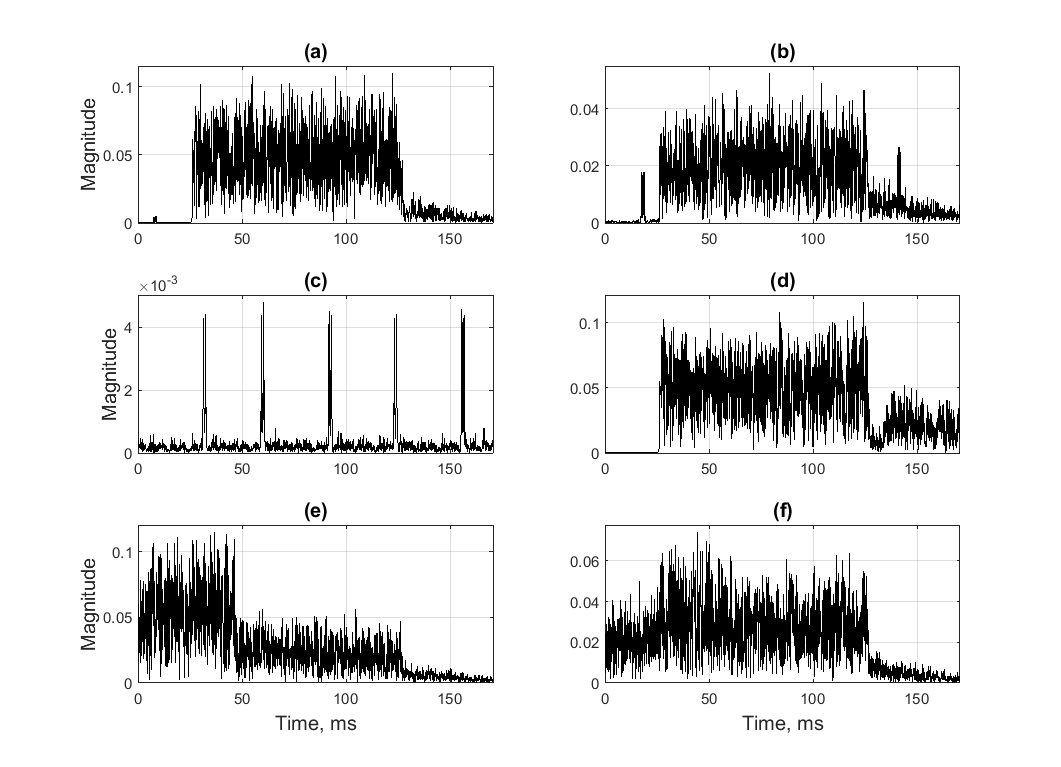}
\caption{Examples of signal segments extracted in the process of the data packet detection.}
\label{Fig:ICMUD-Packet-Detection-Examples}
\end{figure}
Examples of signal segments extracted in the process of the data packet detection are shown in Fig.~\ref{Fig:ICMUD-Packet-Detection-Examples}. 
Fig.~\ref{Fig:ICMUD-Packet-Detection-Examples}(a) shows a typical extracted data packet with almost no interference apart from the multipath interference in the form of the long tail of the packet. Fig.~\ref{Fig:ICMUD-Packet-Detection-Examples}(b) shows an extracted packet accompanied with an impulse interference and more intensive reverberation. Fig.~\ref{Fig:ICMUD-Packet-Detection-Examples}(c) shows a case of false alarm, when a sequence of impulses is recognized as a data packet; this will be eliminated in the further baseband processing. Fig.~\ref{Fig:ICMUD-Packet-Detection-Examples}(d) shows a strong detected packet and a weaker packet that follows and interferes with the tail of the detected packet. In Fig.~\ref{Fig:ICMUD-Packet-Detection-Examples}(e), there can be seen two packets collided; a weak packet has been detected, while an earlier arrived stronger packet introduces the MAI. In Fig.~\ref{Fig:ICMUD-Packet-Detection-Examples}(f), a stronger detected packet is affected by a weaker packet due to the collision.   

Note that the condition~(\ref{Eq:Test-1}) may result in a large number of false alarms, as, for example, is the case demonstrated in Fig.~\ref{Fig:ICMUD-Packet-Detection-Examples}(c). In our lake experiment, we extracted on average about two signal segments per one transmitted packet. This results in approximately double computational load of the baseband processing compared to an ideal packet detection. To reduce the number of false detections, the test~(\ref{Eq:Test-1}) can be replaced with a test $I(u,q) > \eta[I(u,q-3) + I(u,q+3)]$, where the factor $\eta$ can be tuned, or some other tests; we did not investigate such opportunities in this paper.

\subsection{Frequency domain channel estimation \label{Subsec:Channel-Estimation}}

The sparse channel estimation (see more details in~\cite{zakharov2019data}) is implemented in two steps. The first step is to find a support $\Gamma$, i.e. delays of multipath components with significant magnitudes, and we limit the support cardinality by a constant $M$, which in our design is set to $M=30$. The second step is to find multipath complex amplitudes for delays in the support. The support estimation is based on searching for the maxima of the cross-correlation function of the received and transmitted signals as follows.   

Samples of the Doppler corrected baseband signal at the symbol rate $F_d$ in the processing branches are represented as 
\begin{eqnarray}
\tilde{z}_b(n) = \tilde{r}_\gamma (2n + b), \  b = 1 , 2, \ n=0,\ldots,N_\text{FFT}-1 .
\label{Eq:Received-LPE-for-ChEst}
\end{eqnarray}
These samples are transformed into the frequency domain using the FFT: 
\begin{eqnarray}
Z_b(k) = \sum_{n = 0}^{N_\text{FFT}-1} \tilde{z}_b(n) e^{-j 2 \pi k n / N_\text{FFT}} .
\label{Eq:RSignal-Freq-Domain}
\end{eqnarray}

The pilot signal $p(n)$ and tentatively demodulated data symbols $\tilde{d}(n)$, $n = 0, \ldots N-1$, are combined into estimates of the transmitted symbols, $\hat{s}(n) = p(n) + j\tilde{d}(n)$, and transformed into the frequency domain:   
\begin{eqnarray}
\hat{S}(k) = \sum_{n = 0}^{N_\text{FFT}-1} \hat{s}(n) e^{-j 2 \pi k n / N_\text{FFT}}.
\label{Eq:Pilot-Freq-Domain}
\end{eqnarray}
At the first turbo iteration, $\tilde{d}(n) = 0$. At subsequent iterations, $\tilde{d}(n)$ is the hard output of the encoder whose input is the decoded message.   

The cross-correlation of the received and transmitted signals is computed as the inverse FFT: 
\begin{eqnarray}
\gamma_b(n) =   \frac{1}{N_\text{FFT}} \sum_{k = 0}^{N_\text{FFT}-1} Z_b(k)\hat{S}^*(k) e^{j 2 \pi k n / N_\text{FFT}} .
\label{Eq:Cross-Correlation}
\end{eqnarray}
The first up to $M$ maxima of $|\gamma_b(n)|^2$ for $n \in [N_\text{FFT} - \Upsilon, N_\text{FFT}] \cup [0, \Upsilon]$, exceeding a threshold $t_\text{paths} = \kappa \max_n |\gamma_b(n)|^2$, where $0 < \kappa < 1$ and we use $\kappa = 4\cdot 10^{-4}$, determine the support $\Gamma$ of a cardinality $|\Gamma| = M_\text{max} \leq M$. The parameter $\Upsilon$ defines the maximum delay spread $[-\Upsilon/F_d,\Upsilon/F_d]$ in the channel. With $\Upsilon = 150$, used in our design, a delay spread of 50~ms is covered by the estimator.

In the second step, the multipath amplitudes are found using the regularized least squares algorithm as follows. Denote 
\begin{eqnarray}
\mathbf{B}^{(q)} =  \left\{ e^{-j 2 \pi k m(q) / N_\text{FFT}} \right\}_{k = 0}^{N_\text{FFT}-1} , \ \ m(q) \in \Gamma ,
\label{Eq:Sparse-Basis-Functions}
\end{eqnarray}
an $N_\text{FFT} \times 1$ vector, representing the $q$th column of the $N_\text{FFT} \times M_\text{max}$ matrix $\mathbf{B}$, where $m(q)$ represents the channel delay of the $q$th path in the support $\Gamma$. Denote $\mathbf{S}= \text{diag}\{[\hat{S}0),\ldots,\hat{S}(N_\text{FFT}-1)]^T \}$.  The $M_\text{max} \times 1$ vector $\hat{\mathbf{c}}_b$ of multipath amplitudes is found by solving the $M_\text{max} \times M_\text{max}$ system of equations
\begin{eqnarray}
\mathbf{G}_b \hat{\mathbf{c}}_b  = \boldsymbol{\xi}_b  ,
\label{Eq:System-LS}
\end{eqnarray}
where 
\begin{eqnarray}
\mathbf{G}_b  &=& \mathbf{B}^H \mathbf{S}^{H} \mathbf{S} \mathbf{B} + \varepsilon \mathbf{I}, \label{Eq:System-G} \\
\boldsymbol{\xi}_b &=& \mathbf{B}^H \mathbf{S}^{H} \tilde{\mathbf{z}}_b , 
\label{Eq:System-xi}
\end{eqnarray}
$\varepsilon > 0$ is a small number, $\mathbf{I}$ is the identity matrix, and $\tilde{\mathbf{z}}_b$ is a vector with elements $\tilde{z}_b(n)$ in~(\ref{Eq:Received-LPE-for-ChEst}). The channel frequency response is then estimated as:      
\begin{eqnarray}
\hat{\mathbf{h}}_b  = \mathbf{B} \hat{\mathbf{c}}_b .
\label{Eq:Frequency-Response-Estimate}
\end{eqnarray}

The complexity of the channel estimator is mainly due to computation of the matrix $\mathbf{G}_b$ in~(\ref{Eq:System-G}) with elements $G_{p,q}$. This computation can be efficiently done by first computing (by using FFT) the vector
\begin{eqnarray}
\rho(n) =  \sum_{k = 0}^{N_\text{FFT}-1} | P(k)|^2  e^{-j 2 \pi k n / N_\text{FFT}} , 
\label{Eq:FFT-for-rho}
\end{eqnarray}
then elements of $\mathbf{G}_b$ are found as 
\begin{eqnarray}
G_{p,q} = \rho( | m(q) - m(p)| ) . 
\label{Eq:Elements-of-G-via-rho}
\end{eqnarray}
Note that elements of the vector $\boldsymbol{\xi}_b$ in~(\ref{Eq:System-xi}) are elements of $\gamma_b(n)$ at the support $\Gamma$, and they are already available. 

The direct solution of the normal system of equations in~(\ref{Eq:System-LS}) can be done using such a classical method as the Cholesky decomposition. In our design, to reduce the complexity and make the algorithm suitable for implementation on fixed-point DSP and hardware platforms, the DCD (dichotomous coordinate descent) algorithm is used~\cite{Zakharov:DCD, liu2009architecture}.

\subsection{IC-Rake combiner: MRC of multipath signal components \label{Subsec:IC-Rake-combiner}}

In this subsection we use $t$ to denote the time variable, which can be continuous or discrete. In the former case, the spectrum $S(\omega)$ of a signal $s(t)$ is the continuous-time Fourier transform
\begin{eqnarray}
S(\omega) = \int_{-\infty}^{\infty} s(t) e^{-j \omega t} dt .
\label{Eq:ICRake-Fourier-Transform}
\end{eqnarray}
In the later case, the spectrum is the discrete-time Fourier Transform
\begin{eqnarray}
S(\omega) = \sum_{t = -\infty}^{\infty} s(t) e^{-j \omega t} .
\label{Eq:ICRake-DTFT}
\end{eqnarray}

In a multipath channel, the received signal without noise can be represented as 
\begin{eqnarray}
x(t) = \sum_{m = 1}^{M} h_m s(t - \tau_m),
\label{Eq:ICRake-received-signal}
\end{eqnarray}
where $s(t)$ is the transmitted signal, $M$ is the number of multipath arrivals, and $\tau_m$ and $h_m$ are delay and amplitude of the $m$th path, respectively. The classical Rake combiner~\cite{proakis2008digital} produces a combined signal given by 
\begin{eqnarray}
z(t) &=& k \sum_{n = 1}^{M} h_n^* x(t + \tau_n) \label{Eq:ICRake-Rake} \\
&=& s(t) \left( k \sum_{n = 1}^{M} |h_n|^2 \right) \label{Eq:ICRake-Rake-MRC}\\ 
&+& k \sum_{n = 1}^{M} \sum_{m = 1, m \ne n }^{M} h_n^* h_m s(t + \tau_n - \tau_m),
\label{Eq:ICRake-Rake-Interference}
\end{eqnarray}
where $k$ is an arbitrary constant. It performs the MRC of the multipath signal components arriving at the receiver with the delays $\tau_m$, $m=1,\ldots,M$, and represented by its first term~(\ref{Eq:ICRake-Rake-MRC}), and thus it provides the highest signal-to-noise ratio (SNR) at its output. However, such a combining ignores the multipath interference represented by the second term~(\ref{Eq:ICRake-Rake-Interference}). 
The term~(\ref{Eq:ICRake-Rake-MRC}) is an ideally equalized signal, i.e. $z(t) = p(t)$ assuming that the term in~(\ref{Eq:ICRake-Rake-Interference}) is zero, when  
\begin{eqnarray}
k = \left( \sum_{m = 1}^{M} |h_m|^2 \right)^{-1}. 
\label{Eq:ICRake-k}
\end{eqnarray}

In the frequency domain, the received signal~(\ref{Eq:ICRake-received-signal}) is represented as:  
\begin{eqnarray}
X(\omega) = S(\omega) H(\omega),
\label{Eq:ICRake-received-signal-Freq}
\end{eqnarray}
where $S(\omega)$ and $X(\omega)$ are spectra of the transmitted and received signals, respectively, and $H(\omega)$ is the channel frequency response given by 
\begin{eqnarray}
H(\omega) = \sum_{m = 1}^{M} h_m e^{-j\omega \tau_m} .
\label{Eq:ICRake-channel-frequency-response}
\end{eqnarray}
The combining in~(\ref{Eq:ICRake-Rake}) takes the frequency-domain form
\begin{eqnarray}
Z(\omega) = k X(\omega) H^*(\omega).
\label{Eq:ICRake-Rake-MRC-Freq}
\end{eqnarray}
Note that according to Parseval's theorem, assuming that the complex exponentials $e^{-j\omega \tau_m}$, $m = 1, \ldots, M$, are orthogonal over the baseband frequency interval $\omega \in [-\pi F_d, \pi F_d]$, we have 
\begin{eqnarray}
\sum_{m = 1}^{M} |h_m|^2 = \frac{1}{2\pi}\int_{-\pi F_d}^{\pi F_d} |H(\omega)|^2 d \omega = |H(\omega)|^2_\text{av} \ , 
\label{Eq:ICRake-k-Parseval}
\end{eqnarray}
i.e., $k$ in~(\ref{Eq:ICRake-k}) is an inverse of the average square magnitude of the channel frequency response, and~(\ref{Eq:ICRake-Rake-MRC-Freq}) can be rewritten in the form 
\begin{eqnarray}
Z(\omega) =  \frac{H^*(\omega)}{|H(\omega)|^2_\text{av}} X(\omega).
\label{Eq:ICRake-Rake-Freq}
\end{eqnarray}
A linear filter with the frequency response 
\[
H_\text{Rake}(\omega) = \frac{H^*(\omega)}{|H(\omega)|^2_\text{av}}  
\]
represents the Rake combiner.

Note that the zero-forcing linear equalizer often used in high-data-rate systems for recovering signals in multipath channels, since it completely eliminates the ISI, is described by a filter with the frequency response 
\[
H_\text{ZF}(\omega) = \frac{H^*(\omega)}{|H(\omega)|^2}. 
\]
In noisy scenarios, the equalizer performance can be improved by introducing a regularisation factor $\epsilon > 1$, so that the equalizer filter is represented by the frequency response
\[
H_\text{LE}(\omega) = \frac{H^*(\omega)}{|H(\omega)|^2 + \epsilon}.
\]

The received signal in~(\ref{Eq:ICRake-received-signal}) can also be rewritten as
\begin{eqnarray}
x(t) = \sum_{m = 1}^{M} x_m(t) ,
\label{Eq:ICRake-received-signal-components}
\end{eqnarray}
where $x_m(t) = h_m s(t - \tau_m)$ is the signal component arriving at the receiver via the $m$th path. In the frequency domain, we have
\begin{eqnarray}
X(\omega) = \sum_{m = 1}^{M} X_m(\omega) ,
\label{Eq:ICRake-received-signal-components-Freq}
\end{eqnarray}
where $X_m(\omega) = h_m S(\omega) e^{-j\omega \tau_m}$. 

An ideal (i.e., without the ISI) MRC of multipath signal components would perform the following operation: 
\begin{eqnarray}
z_0(t) = k \sum_{n = 1}^{M} h_n^* x_n(t + \tau_n) 
\label{Eq:ICRake-Ideal-MRC}
\end{eqnarray}
or, in the frequency domain,  
\begin{eqnarray}
Z_0(\omega) &=& k \sum_{n=1}^{M} h_n^* e^{j\omega \tau_m} X_n(\omega) .
\label{Eq:ICRake-Ideal-MRC-Freq}
\end{eqnarray}
Our task is to estimate $X_n(\omega)$ and use the combiner~(\ref{Eq:ICRake-Ideal-MRC-Freq}). 

Supposing that, in the receiver, turbo iterations are used that alternate between the channel estimation and data decoding, we can assume that, at the start of a turbo iteration, estimates $\hat{h}_n$ of the channel parameters $h_n$ are available, as well as an estimate $\hat{S}(\omega)$ of the spectrum of the transmitted signal $S(\omega)$. We also assume that the delays $\tau_n$ are known; this is the case, when the channel estimator is implemented in a digital form, thus the delays are fixed and represent the channel taps. In this case, an estimate of $X_n(\omega)$ can be obtained in the form
\begin{eqnarray}
\hat{X}_n(\omega) = X(\omega) - \sum_{m=1, m \neq n}^{M} \hat{h}_m e^{-j\omega \tau_m} \hat{S}(\omega).  
\label{Eq:ICRake-ICRake-Xn-estimate}
\end{eqnarray}
By replacing $X_n(\omega)$ in (\ref{Eq:ICRake-Ideal-MRC-Freq}) with $\hat{X}_n(\omega)$ from~(\ref{Eq:ICRake-ICRake-Xn-estimate}), we obtain
\begin{eqnarray}
\hat{Z}_0(\omega) = k \sum_{n=1}^{M} \hat{h}_n^* e^{j\omega \tau_n} \hat{X}_n(\omega)  \\
            = k \hat{H}^*(\omega) \left[X(\omega) - \hat{S}(\omega)\hat{H}(\omega) \right] 
            + k \hat{S}(\omega) \sum_{m=1}^{M} |\hat{h}_m|^2 , 
\label{Eq:ICRake-Ideal-MRC-Freq-est}
\end{eqnarray}
where $\hat{H}(\omega) = \sum_{m = 1}^{M} \hat{h}_m e^{-j\omega \tau_m}$ is an estimate of the channel frequency response $H(\omega)$. By choosing the constant $k$ according to~(\ref{Eq:ICRake-k}) as
\begin{eqnarray}
k = \left( \sum_{m = 1}^{M} |\hat{h}_m|^2 \right)^{-1}  ,
\label{Eq:ICRake-k-hat}
\end{eqnarray}
we arrive at the frequency-domain combined signal in the form
\begin{eqnarray}
\hat{Z}_0(\omega) = \hat{S}(\omega) + k \hat{H}^*(\omega) \left[X(\omega) - \hat{S}(\omega)\hat{H}(\omega) \right] . 
\label{Eq:ICRake-ICRake-combined-signal-Freq}
\end{eqnarray}
We call this the IC-Rake combiner. 

With no \textit{a priori} information on the transmitted signal, i.e., $\hat{S}(\omega) = 0$, from~(\ref{Eq:ICRake-ICRake-combined-signal-Freq}), we arrive at the classical Rake combiner in~(\ref{Eq:ICRake-Rake-MRC-Freq}). With perfect estimates of the channel, $\hat{H}(\omega) = H(\omega)$, and the signal, $\hat{S}(\omega) = S(\omega)$, we obtain from~(\ref{Eq:ICRake-ICRake-combined-signal-Freq}) the ideal result $\hat{Z}_0(\omega) = S(\omega)$. In general case, the combiner updates the signal estimate obtained at the previous turbo iteration by applying the classical Rake combiner to the residual signal $X(\omega) - \hat{S}(\omega)\hat{H}(\omega)$ resulting from inaccurate estimates of the channel and transmitted signal at the previous turbo iteration.   

In the case of the superimposed data and pilot signals, used in our modem, we have $S(\omega) = P(\omega) + j D(\omega)$. Before the start of turbo iterations, the pilot signal is known, but the data signal is unknown. Therefore, at the first iteration we have $\hat{S}(\omega) = P(\omega)$. At further turbo iterations, we use $\hat{S}(\omega) = P(\omega) + j \hat{D}(\omega)$, where $\hat{D}(\omega)$ is the spectrum of of the data signal estimate obtained at the previous turbo iteration.  

\begin{figure}
\centering
\includegraphics[width=9cm]{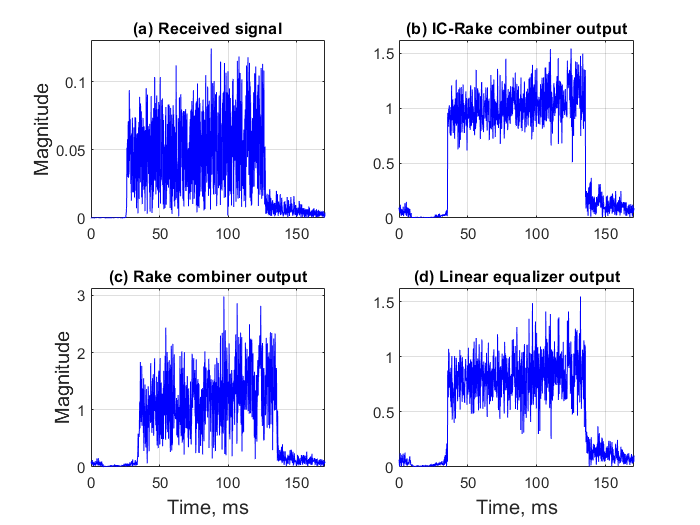}
\caption{Examples of equalized data packets obtained by the IC-Rake combiner, Rake combiner, and linear equalizer.}
\label{Fig:ICMUD-Equalized-Packet-Examples}
\end{figure}
Fig.~\ref{Fig:ICMUD-Equalized-Packet-Examples} shows examples of a received data packet [Fig.~\ref{Fig:ICMUD-Equalized-Packet-Examples}(a)] after applying the IC-Rake combiner [Fig.~\ref{Fig:ICMUD-Equalized-Packet-Examples}(b)], Rake combiner [Fig.~\ref{Fig:ICMUD-Equalized-Packet-Examples}(c)], and linear equalizer [Fig.~\ref{Fig:ICMUD-Equalized-Packet-Examples}(d)]. The important part of the equalized signals are $N=600$ data samples (corresponding to 100~ms time interval) in the centre of the plots; these are extracted for further processing, while the rest part is discarded. It can be seen that the best equalization is provided by the IC-Rake combiner, followed by the linear equalizer, whereas the Rake combiner shows the worst result.

\subsection{Data packet regeneration \label{Subsec:Data-Packet-Recovery}}

\begin{figure}
\centering
\includegraphics[width=8cm]{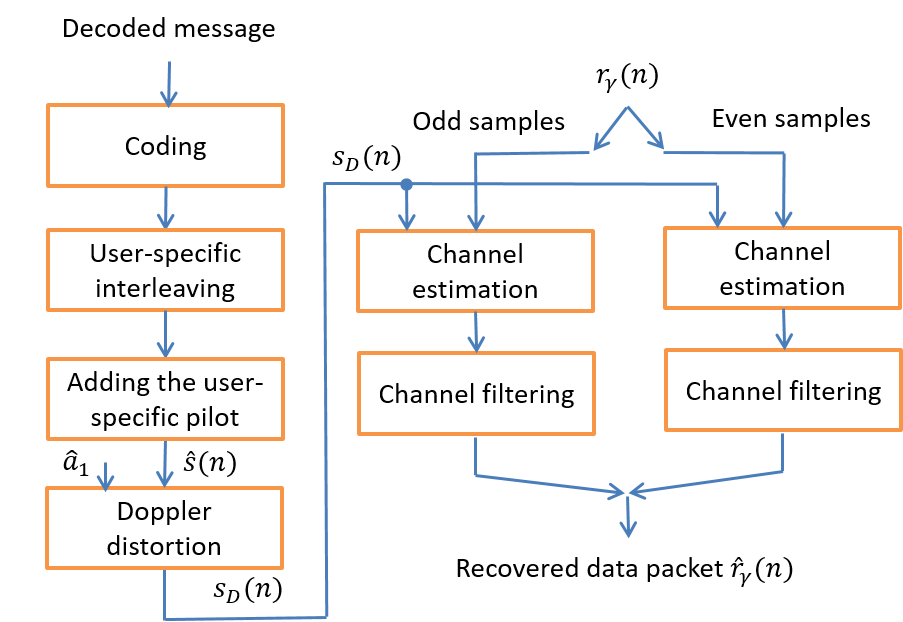}
\caption{Structure of the signal recovery scheme.}
\label{Fig:ICMUD-Signal-Recovery}
\end{figure}

For recovery of a data packet the processing scheme shown in Fig.~\ref{Fig:ICMUD-Signal-Recovery} is used. Assuming that the original (without Doppler correction) extracted signal segment containing the data packet is $r_\gamma(n)$, an estimate of the transmitted data is $\hat{s}(n)$ and Doppler estimate is $\hat{a}_1$, we imitate the signal distortion introduced in the channel as follows. We first introduce the Doppler distortion into the recovered data signal $\hat{s}(n)$: 
\begin{eqnarray}
s_\text{D}(n) = \hat{s}(n) e^{j2\pi \hat{a}_1 (f_c / F_d) n} . 
\label{Eq:Doppler-distorted-data}
\end{eqnarray}
The signal $s_\text{D}(n)$ as a pilot signal and the received signal $r_\gamma(n)$ (i.e., the original extracted signal segment) are used for the channel estimation, which is done using the same estimation algorithm as described in subsection~\ref{Subsec:Channel-Estimation}. The channel frequency response estimate $\hat{H}(\omega)$ and the spectrum $S_\text{D}(\omega)$ of the signal $s_\text{D}(n)$ are used to perform the channel filtering and produce the spectrum $\hat{R}(\omega) = \hat{H}(\omega) S_\text{D}(\omega)$ of the recovered signal, which is then transformed into the time-domain signal $\hat{r}_\gamma(n)$ using the inverse FFT. This is performed separately for odd and even samples of the input signal, producing correspondingly odd and even samples of the recovered data packet.   

To obtain the signal estimate $\hat{s}(n)$, we repeat the processing used in the transmitter, including the coding, user-specific interleaving, and adding the user-specific pilot symbols according to~(\ref{Eq:Superimposed-Symbols}).  

Using first the Doppler distortion and then the channel estimation allows resolving the phase uncertainty, which otherwise would be present in the recovered signal.   
Note that here we introduce the Doppler distortion as a frequency shift without the signal resampling. We have found that at this processing step such a simplified Doppler model is working well for our experimental scenarios with almost static transmitters and receivers. For more dynamic scenarios, the resampling will be required. 

The result of the data packet recovery is a signal segment $\hat{r}_\gamma(n)$ at the sampling rate twice the symbol rate, which is a baseband representation of the data packet at the output of the channel.

\subsection{Interference cancellation \label{Subsec:IC-Data-Packet-Collisions}}

There are several loops of interference cancellation in the proposed receiver as shown in Fig.~\ref{Fig:ICMUD-Receiver}. 
The first loop is the IC of the multipath interference in the IC-Rake combiner as described in subsection~\ref{Subsec:IC-Rake-combiner}; this loop includes $I_1$ turbo iterations, and in our experiments we set $I_1 = 9$. The other two loops, called the packet IC with $I_2$ iterations and signal IC with $I_3$ iterations, are as follows.  
 
For the packet IC, timestamps $\tau_\gamma$ are used to find overlapping signal segments $r_\gamma(n)$ by analysing the timestamp difference $\Delta = \tau_\gamma - \tau_\nu$. If $|\Delta| < 2 N_\text{FFT}$, the segments are overlapped, and $r_\gamma(n)$ is updated as 
\begin{eqnarray}
r_\gamma(n) \leftarrow r_\gamma(n) - \hat{r}_\nu(n - \Delta) 
\label{Eq:packet-IC-update}
\end{eqnarray}
for $n$ within the overlapping part of the two signal segments. This however only performed if the $\text{SNR}_\text{MRC}$ exceeds a predefined threshold, which is in our case is set to 0 dB; $\text{SNR}_\text{MRC}$ is calculated as in~(\ref{Eq:SNR-Estimate}) by replacing $y_b(n)$ with $y(n)$ from~(\ref{Eq:Diversity-Combining-Data}). 

For the signal IC, the recovered signal is subtracted from the baseband signal $r_\text{B}(n)$: 
\begin{eqnarray}
r_\text{B}(n-\tau_\nu) \leftarrow r_\text{B}(n-\tau_\nu) - \hat{r}_\nu(n), n=0,\ldots,2 N_\text{FFT} -1,  
\label{Eq:signal-IC-update}
\end{eqnarray}
This IC is useful since due to the packet collision, there is a high probability of a missed packet detection. After the signal IC, such a probability is reduced. The signal update~(\ref{Eq:signal-IC-update}) is performed only if the CRC check is successful.

\section{Lake experiment \label{Sec:Numerical-Sea}}

\begin{figure}
\centering
\includegraphics[width=8cm]{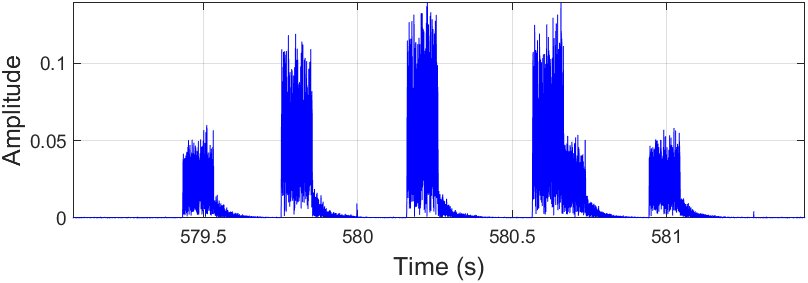}
\includegraphics[width=8cm]{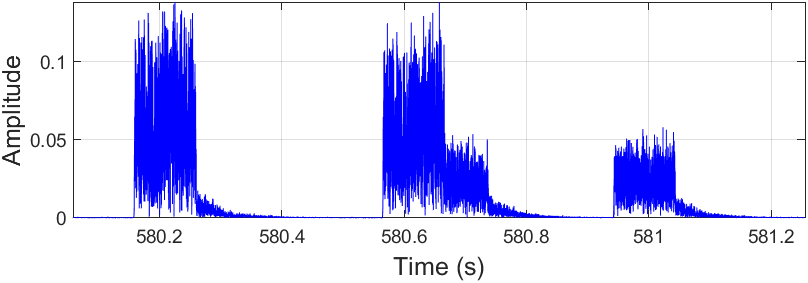}
\caption{Overlapping received data packets at H1.}
\label{Fig:Overlapping-data-packets}
\end{figure}

\begin{figure}
\centering
\includegraphics[width=6cm]{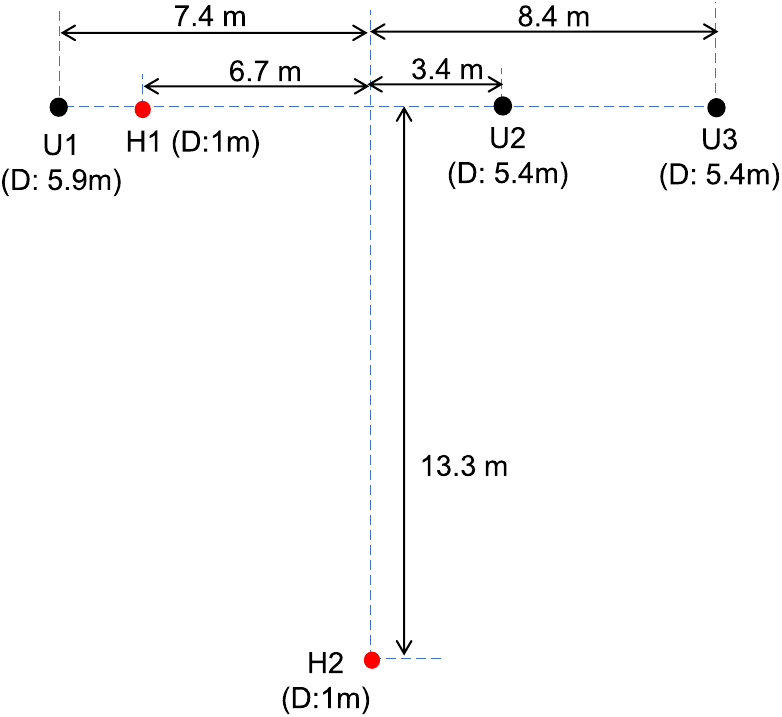}
\caption{Positions of the user transmitters (U1, U2, U3) and hydrophones (H1, H2) during the lake experiment; parameter D indicates the depth of the transmitter or receiver.}
\label{Fig:TxRx_positions}
\end{figure}
\begin{figure}
\centering
\includegraphics[width=8cm]{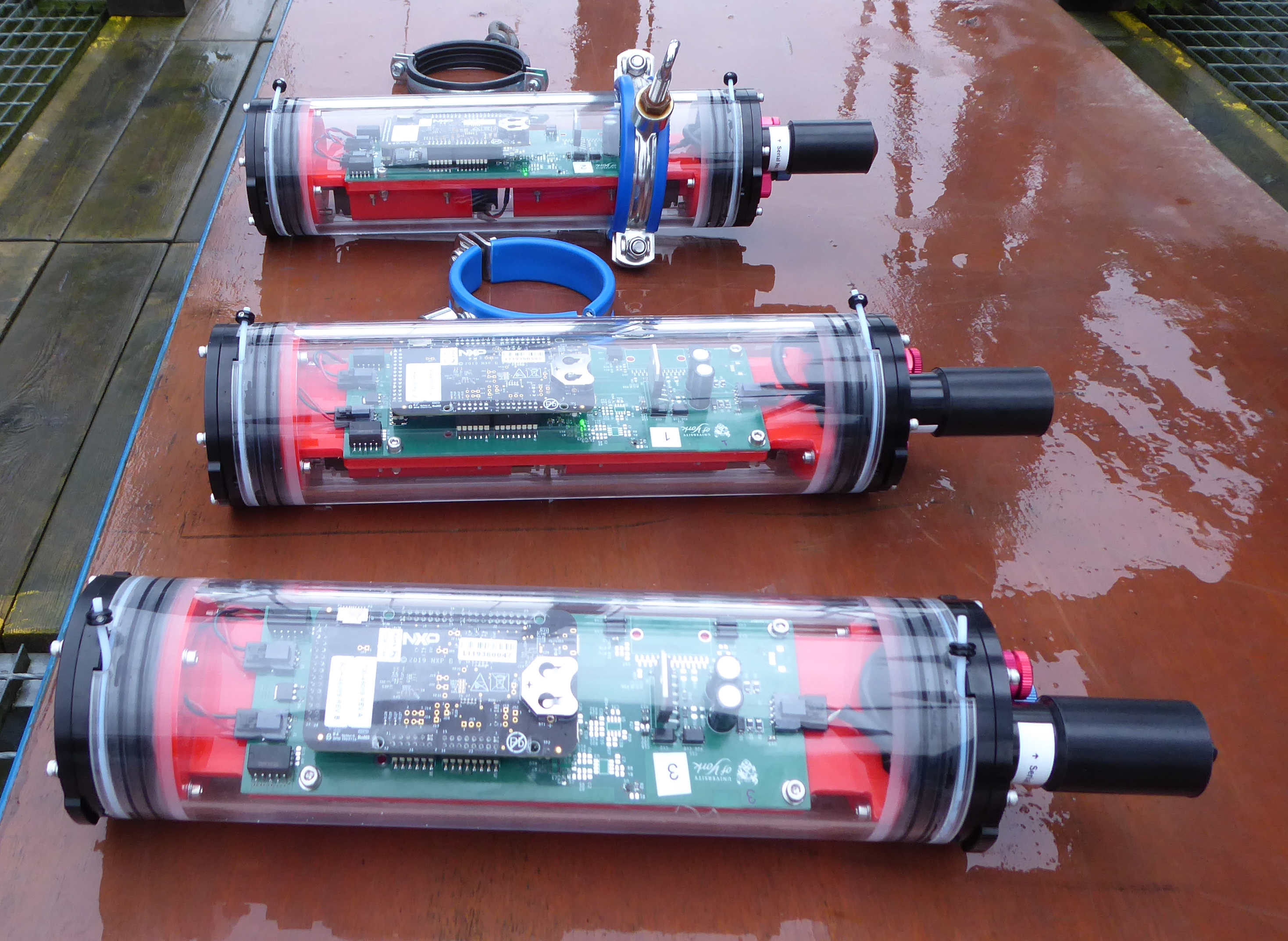}
\caption{Transmitters used in the lake experiment.}
\label{Fig:Three-Tx}
\end{figure}

\begin{table}[]
\caption{Direct distances between transmitters  and hydrophones. \label{Table:TxRx_Distances}}
\begin{center}
\begin{tabular}{lccc}
\hline 
                      &   U1  &   U2  &  U3   \\
 \hline  
  H1 &  4.9~m   &  11.0~m  &  15.7~m \\
 \hline  
  H2 &  16.0~m   &  14.4~m  &  16.3~m \\
 \hline  
\end{tabular}
\end{center}
\end{table}
The experiment was conducted in a lake of depth about 7~m. Fig.~\ref{Fig:TxRx_positions} shows positions of transmitters and receivers during the experiment. Fig.~\ref{Fig:Three-Tx} shows three transmitters used in the experiment. Table~\ref{Table:TxRx_Distances} shows the direct distances between transmitters and receivers. 
During 20~min of the experiment, three users simultaneously and asynchronously transmitted data packets with parameters shown in Table~\ref{Table:Data_packet}. Every user transmitted a burst of 3~data packets every 8~s. Time patterns of the 8~s period were as following: (i) user~1 (U1) transmitted a burst of 3 packets with the packet period 0.4~s; (ii) user~2 (U2) transmitted with the packet period 1.2~s; and (iii) user~3 (U3) transmitted with the packet period 2.8~s. In total, the three users transmitted 2100 data packets. The time clocks at the transmitters were set with an accuracy of about 1$\%$, and therefore, during the experiment, relative time positions of user packets varied resulting in multiple packet collisions.  

\begin{figure}
\centering
\includegraphics[width=8cm]{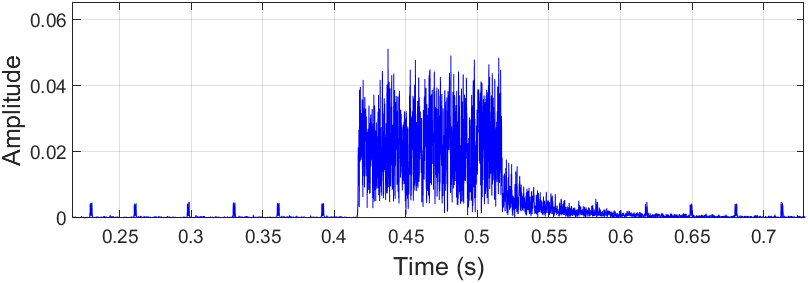}
\includegraphics[width=8cm]{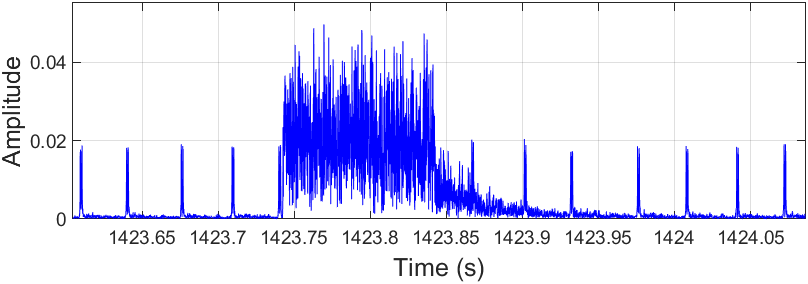}
\caption{Impulse interference in the baseband signal of hydrophone~H1 at the beginning (top plot) and in the middle (bottom plot) of the 20-min experiment.}
\label{Fig:Impulse-interference}
\end{figure}
The level of the transmitted signals was high enough to ignore the ambient noise. However, the communication most of the time was affected by impulse interference due to other (unknown) acoustic systems operating in the vicinity of the experimental area. Fig.~\ref{Fig:Impulse-interference} illustrates the impulse interference in comparison to the level of a received data packet. This interference however did not seem to affect the receiver performance. 

\begin{figure}
\centering
\includegraphics[width=9cm]{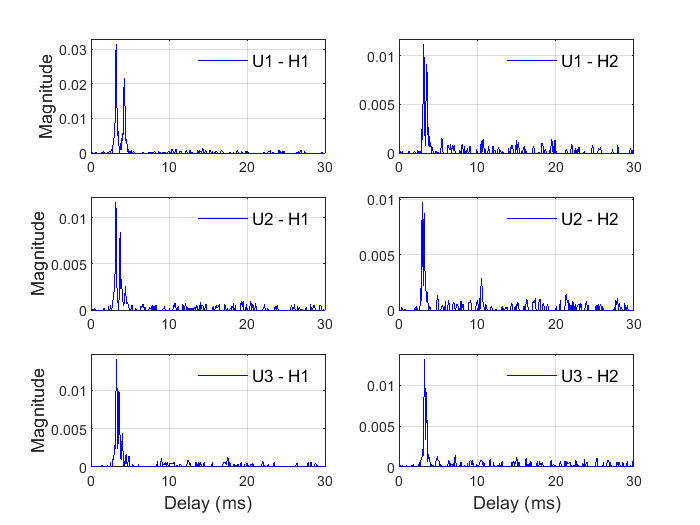}
\caption{Impulse responses of UWA channels between all transmitters (U1, U2, and U3) and receivers (H1 and H2).}
\label{Fig:Impulse_responses}
\end{figure}
Fig.~\ref{Fig:Impulse-interference} also illustrates a long reverberation tail indicating a long multipath delay profile of the channels. Fig.~\ref{Fig:Impulse_responses} shows snapshots of magnitudes of the channel impulse responses between all transmitters and receivers participating in the experiment, for the first 30~ms of the channel delays. It can be seen that apart from several strong paths with small delays, there are multiple weaker paths with longer delays, which are responsible for the reverberation tails of received data packets as is observed in~Fig.~\ref{Fig:Impulse-interference}. It can also be seen in~Fig.~\ref{Fig:Impulse_responses} that the receiver H2 is characterised by stronger reverberation than the receiver H1, and therefore the H2 performance is expected to be somewhat inferior to that of~H1. 

Despite the relatively short distances between the transmitters and receivers, the multipath channels are quite challenging for the communication due to the long channel delay profiles, as can be seen in  Fig.~\ref{Fig:Impulse-interference} and Fig.~\ref{Fig:Impulse_responses}. 
It is seen in~Table~\ref{Table:TxRx_Distances} that the U1-H1 distance is much smaller than U2-H1 and U3-H1 distances, thus this link has the lowest signal attenuation, as can also be seen in~Fig.~\ref{Fig:Impulse_responses}, and therefore SNR for the packets arrived from U1 to H1 should be the highest. Powers of user signals arrived at H1 are clearly different, whereas all user signals arrived at H2 are of about the same power. Thus, the signal at H2 is expected to be more challenging for the IC than the signal at~H1.      

Note that many collisions with partially overlapped packets are successfully resolved without the IC cancellation due to the data coding. If packets are significantly overlapped, the packet with a higher power might still be successfully decoded, while the weaker packet will be lost without the IC.  

We now consider the receiver performance in a number of scenarios. The performance is measured by the frame error rate (FER) computed as a ratio of the number of lost packets to the total number of transmitted packets. 

\subsection{Network with channel utilization~11\%}

Since there are 3 packets transmitted by every user within a period of 8~s and the packet duration is 0.1~s, the channel utilization is $(3 \times 3 \times 0.1) / 8 \approx 11\%$. In this subsection, we investigate the receiver performance at two positions, H1 and H2, without and with extra noise added to the received signal.

\subsubsection{Reception at H1 and H2 without extra noise}
\begin{table}[]
\caption{FER at H1. 
\label{Table:Ch1-no-noise}}
\begin{center}
\begin{tabular}{lccc}
\hline 
                      &   $I_3 = 1$  &   $I_3 = 2$  &   $I_3 = 3$ \\
 \hline  
  $I_2 = 1$ &  $4.6\%$   &  $0.19\%$  &  $0.19\%$ \\
 \hline  
  $I_2 = 2$ &  $0.43\%$   &  $0\%$  &  $0\%$ \\
 \hline  
  $I_2 = 3$ &  $0.43\%$   &  $0\%$  &  $0\%$ \\
 \hline          
\end{tabular}
\end{center}
\end{table}
Results of processing the signal received at H1 are shown in Table~\ref{Table:Ch1-no-noise}. It is seen that without the IC, 4.6\% of the packets (97 packets) are not decoded. These packets are mainly from U2 with the weakest received signal, so the packets are lost when collided with the stronger user packets, and some packets from U3 are also lost during the collisions with much stronger U1.

\begin{figure}
\centering
\includegraphics[width=8cm]{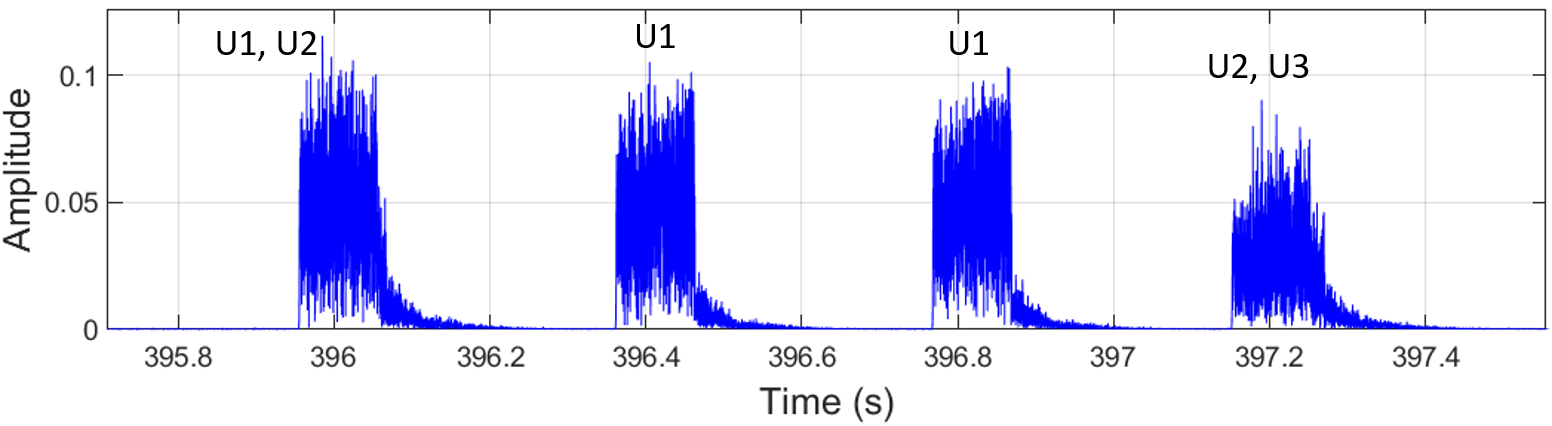}
\caption{For H1, the collision at 397.2~s between U2 and U3 of comparable received signal power is not resolved in the case of $I_3=2$ and $I_2=1$. The collision at 396~s between U1 and U2 is resolved in the case of $I_3=2$ and $I_2=1$, since the signal power of U1 is significantly higher than that of U2.}
\label{Fig:Collisions-397s}
\end{figure}
For the cases when simultaneously the number of packet IC iterations $I_2>1$ and the number of signal IC iterations $I_3>1$, all packets are successfully decoded. However, when $I_3=2,3$ and $I_2=1$, there is a loss of 4 packets (FER = 0.19\%) in two collisions. Fig.~\ref{Fig:Collisions-397s} illustrates a collision between U2 and U3, when both the packets are lost. This happens since both the signals are of about equal power, and the packets are almost completely overlapped, thus none of them were decoded, so the interference cannot be cancelled in the signal IC iterations. In Fig.~\ref{Fig:Collisions-397s}, it can also be seen a collision of U2 packet with a stronger U1 packet. The later was successfully decoded at the first packet IC iteration and subtracted from the received signal at the second iteration, at which the packet U2 was also successfully decoded, thus no packet loss was observed. The use of packet IC iterations alone ($I_3=1$) also does not allow lossless reception. This is due to missing packets during the packet detection. However, with simultaneously $I_2>1$ and $I_3>1$, all packets are successfully decoded.

\begin{table}[]
\caption{FER at H2. 
\label{Table:Ch2-no-noise}}
\begin{center}
\begin{tabular}{lccc}
\hline 
                      &   $I_3 = 1$  &   $I_3 = 2$  &   $I_3 = 3$ \\
 \hline  
  $I_2 = 1$ &  $4.6\%$   &  $1.7\%$  &  $1.7\%$ \\
 \hline  
  $I_2 = 2$ &  $0.57\%$   &  $0.24\%$  &  $0.24\%$ \\
 \hline  
  $I_2 = 3$ &  $0.29\%$   &  $0.24\%$  &  $0.24\%$ \\
 \hline          
\end{tabular}
\end{center}
\end{table}
Table~\ref{Table:Ch2-no-noise} shows the receiver performance for the signal at~H2. At this receiver, as can be seen in Fig.~\ref{Fig:Impulse_responses}, powers of all the user signals are about the same. This creates a problem of detecting significantly overlapped packets, since the code is not strong enough for decoding one of the packets, thus making the IC less efficient. We can see from Table~\ref{Table:Ch2-no-noise} that this is indeed the case, and even with simultaneously $I_3>1$ and $I_2 > 1$, there are still 5 lost packets (FER = 0.24\%). However, this is still a significant improvement compared to the case $I_2=1$ and $I_3=1$, when 96 packets are lost (FER = 4.6\%).  
  
For both the signals, received at H1 and H2, the use of IC iterations over detected packets or over the signal alone does not allow the best performance, whereas when combining such iterations (i.e., $I_2 > 1$ and $I_3 > 1$), a lossless reception at H1 and losing only 5 packets at H2 become possible.

\subsubsection{Reception at H1 and H2 with extra noise}
We now add an extra noise to the received signal to investigate the receiver performance in noisy scenarios. The (white) noise is required to be added to the passband received signal to guarantee the proper noise correlation in the two branches of the baseband processing.  

\begin{table}[]
\caption{FER at H1 with extra noise: SNR$\approx 15$~dB for U1; SNR$\approx 7$~dB for U2; and SNR$\approx 8$~dB for U3. 
\label{Table:Ch1-noise-001}}
\begin{center}
\begin{tabular}{lccc}
\hline 
                      &   $I_3 = 1$  &   $I_3 = 2$  &   $I_3 = 3$ \\
 \hline  
  $I_2 = 1$ &  $5.6\%$   &  $1.1\%$  &  $1.1\%$ \\
 \hline  
  $I_2 = 2$ &  $1.0\%$   &  $0.29\%$  &  $0.29\%$ \\
 \hline  
  $I_2 = 3$ &  $0.86\%$   &  $0.24\%$  &  $0.24\%$ \\
 \hline          
\end{tabular}
\end{center}
\end{table}
Table~\ref{Table:Ch1-noise-001} shows the performance at H1 with a noise level that results in an SNR of about 15~dB for U1, 7~dB for U2 and 8~dB for U3. It can still be seen that the IC significantly reduces the packet loss.  With $I_2=3$ and $I_3>1$, only 5 packets are lost (FER = 0.24\%). This is significantly lower than in the receiver without the IC: 118 lost packets (FER = 5.6\%). 

\begin{figure}
\centering
\includegraphics[width=8cm]{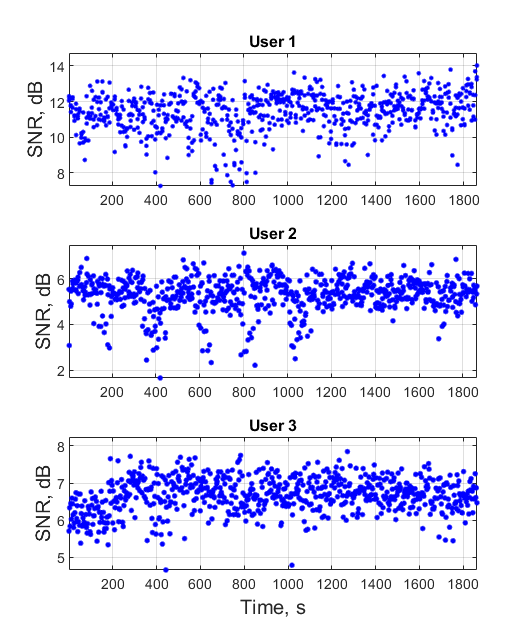}
\caption{For H1, the SNR after the MRC combining in the receiver, measured for data packets decoded without errors. }
\label{Fig:SNR_MRC_v76_1}
\end{figure}
Fig.~\ref{Fig:SNR_MRC_v76_1} shows the SNR$_\text{MRC}$ measured in the receiver without the IC, after the IC-Rake combining. These results are only for data packets that passed CRC tests. It can be seen that these SNR measurements are mostly close to but somewhat lower than SNRs indicated above; this is due to channel/Doppler estimation and equalization errors in the receiver, including the limited number of multipath components involved in the estimation and equalization (at most $M = 30$ multipath components in our experiments). The SNR measurements with significantly lower SNR values most likely correspond to overlapped (but still decoded) packets.   

\begin{table}[]
\caption{FER at H2 with extra noise: SNR$\approx 7$~dB for U1; SNR$\approx 5$~dB for U2; and SNR$\approx 7$~dB for U3.  
\label{Table:Ch2-noise-001}}
\begin{center}
\begin{tabular}{lccc}
\hline 
                      &   $I_3 = 1$  &   $I_3 = 2$  &   $I_3 = 3$ \\
 \hline  
  $I_2 = 1$ &  $7.3\%$   &  $5.2\%$  &  $5.2\%$ \\
 \hline  
  $I_2 = 2$ &  $3.7\%$   &  $3.2\%$  &  $3.2\%$ \\
 \hline  
  $I_2 = 3$ &  $2.8\%$   &  $2.7\%$  &  $2.7\%$ \\
 \hline          
\end{tabular}
\end{center}
\end{table}
Table~\ref{Table:Ch2-noise-001} shows the performance at H2 with the same noise level as above, which now results in an SNR of about 7~dB for U1, 5~dB for U2 and 7~dB for U3. Due to lower SNRs compared to H1, without the IC, FER = 7.3\%, while with the IC, FER = 2.7\%, which is still a valuable improvement. 

\subsection{Network with channel utilization~23\%} 

We now sum together signals received at H1 and H2, with a delay of the H2 signal by 200~ms to prevent overlapping two packets from the same user in the mixed signal. This is equivalent to twice increasing the packet transmission rates by all the users. The total number of packets in the combined signal is 4199 (not $2100 \times 2 = 4200$, since one packet is not present in the combined signal due to the delay). Thus, the channel utilization in the mixed signal is $(6 \times 3 \times 0.1) / 8 \approx 23\%$. In this subsection, we present results of investigation of the receiver performance for the mixed signal without and with an extra noise added to the signal. 

\subsubsection{Reception without extra noise} 
\begin{table}[]
\caption{FER for the mixed signal.  \label{Table:Mix-no-noise}}
\begin{center}
\begin{tabular}{lccc}
\hline 
                      &   $I_3 = 1$  &   $I_3 = 2$  &   $I_3 = 3$ \\
 \hline  
  $I_2 = 1$ &  $33\%$   &  $3.8\%$ &  $2.1\%$ \\
 \hline  
  $I_2 = 2$ &  $28\%$   &  $0.81\%$  &  $0.29\%$ \\
 \hline  
  $I_2 = 3$ &  $28\%$   &  $0.48\%$  &  $0.26\%$ \\
 \hline          
\end{tabular}
\end{center}
\end{table}
In Table~\ref{Table:Mix-no-noise}, FER is shown for the mixed signal. It can be seen that with $I_2=3$ and $I_3=3$, the IC performance is still good, only 11 of 4199 packets are lost (FER~=~0.26\%): 1~packet from~U1, 7~packets from~U2, and 3~packets from~U3. However, without the IC, 33\% of packets are lost. Thus, we observe a significant improvement in the receiver performance when using the IC techniques.


\begin{table}[]
\caption{FER for the mixed signal. Comparison of the performance of the receiver with the Rake combiner, linear equalizer and IC-Rake combiner.   
 \label{Table:Mix-no-noise-Lin-Rake-IC}}
\begin{center}
\begin{tabular}{lccc}
\hline 
                      &  Rake   &   Linear &   IC-Rake  \\
 \hline  
  $I_2 = 1$,  $I_3 = 1$ &  $40\%$   &  $35\%$ &  $33\%$ \\
 \hline  
  $I_2 = 3$, $I_3 = 3$ &  $15\%$   &  $9.1\%$  &  $0.26\%$ \\
 \hline  
\end{tabular}
\end{center}
\end{table}
In Table~\ref{Table:Mix-no-noise-Lin-Rake-IC}, FER is shown for the IC-Rake combiner, linear equalizer and Rake combiner. It can be seen that the IC-Rake combiner outperforms the other techniques and the performance improvement is especially substantial when the IC is applied.

\subsubsection{Reception with extra noise} 
\begin{table}[]
\caption{FER for the mixed signal at H1 and H2 with extra noise 6~dB lower than that for results in  Table~\ref{Table:Ch1-noise-001} and Table~\ref{Table:Ch2-noise-001}.  
\label{Table:Mix-noise-0005}}
\begin{center}
\begin{tabular}{lccc}
\hline 
                      &   $I_3 = 1$  &   $I_3 = 2$  &   $I_3 = 3$ \\
 \hline  
  $I_2 = 1$ &  $34\%$   &  $4.9\%$ &  $3.3\%$ \\
 \hline  
  $I_2 = 2$ &  $29\%$   &  $1.4\%$  &  $0.9\%$ \\
 \hline  
  $I_2 = 3$ &  $29\%$   &  $0.9\%$  &  $0.6\%$ \\
 \hline          
\end{tabular}
\end{center}
\end{table}
In Table~\ref{Table:Mix-noise-0005}, FER is shown for the mixed signal with the extra added noise 6~dB lower in power compared to the noise used to obtain results in Table~\ref{Table:Ch1-noise-001} and Table~\ref{Table:Ch2-noise-001}. Although now 26~packets are lost (FER~=~0.6\%), specifically 4, 16, and 6 packets for U1, U2, and U3, respectively, this is still significantly better compared to 34\% (1424) packets lost without the IC.


\begin{table}[]
\caption{FER for the mixed signal with extra noise 6~dB lower than that for results in  Table~\ref{Table:Ch1-noise-001} and Table~\ref{Table:Ch2-noise-001}. Comparison of the performance of the receiver with the Rake combiner, linear equalizer and IC-Rake combiner. \label{Table:Mix-noise-0005-Lin-Rake-IC}}
\begin{center}
\begin{tabular}{lccc}
\hline 
                      &  Rake   &   Linear &   IC-Rake  \\
 \hline  
  $I_2 = 1$,  $I_3 = 1$ &  $41\%$   &  $35\%$ &  $33\%$ \\
 \hline  
  $I_2 = 3$, $I_3 = 3$ &  $19\%$   &  $13\%$  &  $0.6\%$ \\
 \hline  
\end{tabular}
\end{center}
\end{table}
In Table~\ref{Table:Mix-noise-0005-Lin-Rake-IC}, FER is shown for the IC-Rake combiner, linear equalizer and Rake combiner. With the extra noise, again the IC-Rake combiner is significantly more efficient than the other two techniques.

\begin{table}[]
\caption{FER for the mixed signal with extra noise the same as that for results in  Table~\ref{Table:Ch1-noise-001} and Table~\ref{Table:Ch2-noise-001}.  
\label{Table:Mix-noise-001}}
\begin{center}
\begin{tabular}{lccc}
\hline 
                      &   $I_3 = 1$  &   $I_3 = 2$  &   $I_3 = 3$ \\
 \hline  
  $I_2 = 1$ &  $35\%$   &  $30\%$ &  $30\%$ \\
 \hline  
  $I_2 = 2$ &  $7.8\%$   &  $4.0\%$  &  $3.3\%$ \\
 \hline  
  $I_2 = 3$ &  $6.4\%$   &  $3.4\%$  &  $3.0\%$ \\
 \hline          
\end{tabular}
\end{center}
\end{table}
In Table~\ref{Table:Mix-noise-001}, FER is shown for the mixed signal with the extra added noise of the same power as that to obtain results in  Table~\ref{Table:Ch1-noise-001} and Table~\ref{Table:Ch2-noise-001}. In this case, without the IC, 35\% of packets are lost, whereas with the IC, only 3\% of packets are lost, which is a significant improvement.

\section{Conclusions \label{Sec:Conclusions}}

We have proposed a modem capable of resolving data packet collisions in uplink of random access UWA networks at the Physical Layer by exploiting IC techniques. We have investigated its performance in a lake experiment in 3-user scenarios with a relatively high channel utilization (up to 23\%). Results of the experiment demonstrate that the proposed IC approach is efficient for building UWA networks. 

We have also proposed an IC-Rake combiner that exploits the benefits of turbo iterations in the receiver to provide MRC combining of multipath signal components. This technique resolves the problems of the Rake combiner, which provides a high SNR but suffers from the multipath interference, and linear equalizer, which can ideally provide interference-free signal, but suffers from the noise enhancement. The IC-Rake combiner significantly outperforms the Rake combiner and linear equalizer in the proposed modem.

\section*{Acknowledgement}
This work was supported in part by the U.K. EPSRC through Grants EP/P017975/1 and EP/R003297/1.

\bibliographystyle{IEEEbib}
\bibliography{Autocorrelation_Method,mud_UWA}

\end{document}